\documentclass[aps,pra,reprint,twocolumn,amsmath,amssymb,showpacs,superscriptaddress,longbibliography]{revtex4-2}

\usepackage[T1]{fontenc}
\usepackage[utf8]{inputenc}
\usepackage{amsmath,amssymb,amsfonts,bm,mathtools}
\usepackage{mathrsfs}
\usepackage{graphicx}
\usepackage{dcolumn}
\usepackage{booktabs}
\usepackage{array}
\usepackage{multirow}
\usepackage{tensor}
\usepackage{float}
\usepackage{gensymb}
\usepackage{physics}
\usepackage{xcolor}
\usepackage[mathlines]{lineno}
\usepackage{hyperref}

\hypersetup{
    colorlinks,
    linkcolor={red!50!black},
    citecolor={blue!50!black},
    urlcolor={blue!80!black}
}

\graphicspath{{./}{figures/}}

\newcommand{\epsp}{\epsilon}
\newcommand{\Tzero}{\mathrm{T0}}
\newcommand{\Lone}{\mathrm{L1}}
\newcommand{\Ttwo}{\mathrm{T2}}
\newcommand{\Stwo}{\mathrm{S2}}
\newcommand{\Vtwo}{\mathrm{V2}}

\newcommand{\Rco}{\mathcal{R}}
\newcommand{\fLG}{f_{\mathrm{LG}}}
\newcommand{\chiz}{\chi_z}
\newcommand{\kappaz}{\kappa_z}
\newcommand{\Kz}{\mathcal{K}_z}
\newcommand{\Atrans}{\mathcal{A}}
\newcommand{\psilp}{\psi_{\ell p}}
\newcommand{\psilpS}{\psi_{\ell p}^{\mathrm{S2}}}
\newcommand{\deltaS}{\delta_{\ell p}^{\mathrm{S2}}}

\begin{document}

\title{Transversality Locks Longitudinal Gradients in Structured-Light Quadrupole Transitions}

\author{Kayn A. Forbes}
\email{K.Forbes@uea.ac.uk}
\affiliation{School of Chemistry, University of East Anglia, Norwich Research Park, Norwich NR4 7TJ, United Kingdom}

\author{Dale Green}
\affiliation{School of Engineering, Mathematics, and Physics, University of East Anglia, Norwich Research Park, Norwich NR4 7TJ, United Kingdom}

\author{Abdullah F. Alharbi}
\affiliation{King Abdulaziz City for Science and Technology (KACST), P.O. Box 6086, Riyadh 11442, Saudi Arabia}

\author{A. Salam}
\email{salama@wfu.edu}
\affiliation{Department of Chemistry, Wake Forest University, Winston-Salem, North Carolina 27109, USA}

\begin{abstract}
Electric quadrupole absorption is driven by optical field gradients, not by field amplitudes alone. This distinction is crucial for structured light, where a paraxially small longitudinal field can generate a leading-order longitudinal gradient. For fully vectorial Laguerre-Gaussian modes, Maxwell transversality locks this longitudinal gradient to the transverse structured gradient of the leading paraxial field. This locking is therefore a general electromagnetic constraint, not a focusing-dependent correction that can be removed independently. Decomposing the symmetric-traceless optical field-gradient tensor into spherical components reveals a strongly channel-selective response: the \(\Delta m=0\) transition requires the longitudinal gradient at the retained order, the \(\Delta m=\pm1\) channels receive scalar and vectorial nonparaxial corrections on top of the carrier-scale longitudinal derivative of the transverse field, and the \(\Delta m=\pm2\) channels remain transverse at the same order. These results show that longitudinal optical structure can enter structured-light quadrupole transitions as an indispensable leading-gradient contribution, even when the longitudinal field amplitude itself is paraxially small.

\end{abstract}

\maketitle
\onecolumngrid

\section{Introduction}\label{sec:intro}

Electric quadrupole transitions provide access to light-matter interaction channels beyond the electric-dipole approximation \cite{craig1998molecular, barron2009molecular, Chwalla2009, germann2014observation}. In structured light \cite{forbes2021structured} these transitions are especially important because the electric-quadrupole coupling depends on optical field gradients, allowing spatial phase, amplitude, and polarization structure to enter directly into
internal atomic or molecular excitation amplitudes. The possibility that optical orbital angular momentum (OAM) can influence internal quadrupole transitions was identified in early theoretical work on twisted-light interactions with matter \cite{babiker2002orbital} and subsequently demonstrated in quadrupole excitation of trapped ions \cite{schmiegelow2016transfer}. Related effects have been explored in vortex-driven atomic transitions \cite{schmiegelow2012light, lembessis2013enhanced, afanasev2018experimental, babiker2019atoms, bougouffa2021quadrupole,  alharbi2023significance,mashhadi2024quadrupole}, higher multipole transitions \cite{lange2022excitation}, and chiral molecular or nanophotonic systems \cite{sakai2018nanofocusing, sakai2022plasmonic} where electric-quadrupole couplings can contribute to optical activity \cite{forbes2018optical,forbes2026twisted,forbes2026vortex, lalaguna2025optical}.

For a plane wave,
\begin{equation}
E_i(\mathbf r)=E_{0,i}e^{i\mathbf k\cdot\mathbf r},
\qquad
\mathbf k\cdot\mathbf E_0=0,
\end{equation}
the optical gradient is
\begin{equation}
\nabla_jE_i(\mathbf r)=ik_jE_i(\mathbf r).
\end{equation}
Thus the electric-quadrupole interaction samples the longitudinal propagation derivative of a transverse field. In a real electromagnetic field, e.g. a laser beam, the situation is more subtle. The transverse field varies across the beam profile, producing transverse gradients, while Maxwell transversality also generates a longitudinal field \cite{lax1975maxwell, adams2018optics, forbes2021relevance, forbes2026vortex}. The usual statement that the longitudinal field amplitude is small is therefore not sufficient for quadrupole absorption, because the
interaction depends on gradients rather than on field amplitudes alone.

This distinction is clearest at the level of paraxial ordering. For a
structured beam of waist \(w_0\),
\begin{equation}
\mathbf E^{\Lone}\sim (kw_0)^{-1}\mathbf E^{\Tzero},
\end{equation}
so the first-order longitudinal field (L1) is indeed smaller than the leading transverse field (T0). However, the gradients that drive electric-quadrupole absorption scale as
\begin{equation}
\partial_z\mathbf E^{\Lone}
\sim
\nabla_\perp\mathbf E^{\Tzero}
\sim
w_0^{-1}\mathbf E^{\Tzero}.
\end{equation}
The longitudinal derivative recovers one power of the optical wavenumber, whereas the transverse structured gradient is set by the beam width. Thus the longitudinal field amplitude is paraxially small, but its longitudinal gradient is not smaller than the transverse structured gradient entering the same quadrupole interaction.

The central result of this work is that these two gradients are not merely of the same order: they are locked by Maxwell transversality and must be retained together. Neglecting the longitudinal field therefore removes a leading structured-light contribution to the electric-quadrupole gradient, even when the field amplitude itself is paraxially small. This provides a simple reason why standard transverse-field intuition can fail for structured-light-induced quadrupole transitions. The locking relation itself is not peculiar to orbital angular momentum or to Laguerre-Gaussian modes; it is a direct consequence of Maxwell transversality in any source-free optical field. Structured beams make the constraint operationally important because their transverse fields carry spatial gradients that enter electric-quadrupole absorption.

The effect is also selective. As shown below, the longitudinal gradient is essential for the channel associated with \(\Delta m=0\), absent from the \(\Delta m=\pm2\) channels at the retained order, and enters the \(\Delta m=\pm1\) channels as a weaker vectorial correction to the dominant carrier-scale longitudinal derivative of the transverse field. The role of the longitudinal field is therefore not a generic intensity correction, but a
specific consequence of how structured gradients enter the electric-quadrupole tensor.

We first analyse the classical vectorial Laguerre-Gaussian mode and its optical field-gradient tensor, keeping the spatial tensor algebra separate
from quantum normalization. We then restore the molecular-QED normalization to obtain the electric-quadrupole absorption probability. A probability
retained consistently through second order requires all contributions to each complete spherical component of the gradient tensor to be combined at the amplitude level before its modulus squared is formed. This treatment places the longitudinal and transverse field-gradient contributions on equal footing and establishes which terms modify the leading absorption probability through interference. It thereby provides a consistent framework for determining when longitudinal optical structure is a leading part of structured-light-induced quadrupole absorption rather than a dispensable higher-order correction.

\section{Paraxial ordering of fields and gradients}\label{sec:ordering}

The expansion used here is the standard Lax-type paraxial expansion of a
Maxwell field in powers of the small parameter
\begin{equation}
\epsp \sim \frac{1}{kw_0},
\end{equation}
where \(w_0\) is the beam waist \cite{lax1975maxwell,adams2018optics}. The
leading solution is the zeroth-order transverse paraxial field \(\mathbf E^{\Tzero}\). Maxwell transversality then generates a first-order longitudinal field \(\mathbf E^{\Lone}\) one order smaller in field amplitude. At second order there are two distinct transverse contributions: a co-handed scalar-envelope correction \(\mathbf E^{\Stwo}\) and a cross-handed vectorial Maxwell correction \(\mathbf E^{\Vtwo}\). The superscripts therefore denote the paraxial order of the field amplitude, not directly the order of every field gradient.

The complete retained field has the schematic expansion
\begin{equation}
\mathbf E
=
\mathbf E^{\Tzero}
+
\mathbf E^{\Lone}
+
\mathbf E^{\Stwo}
+
\mathbf E^{\Vtwo}
+
\cdots,
\end{equation}
with
\begin{equation}
\mathbf E^{\Tzero}\sim O(1),
\qquad
\mathbf E^{\Lone}\sim O(\epsp),
\qquad
\mathbf E^{\Stwo}\sim
\mathbf E^{\Vtwo}\sim O(\epsp^2).
\end{equation}
Equivalently, the complete second-order transverse field is
\(\mathbf E^{\Ttwo}=\mathbf E^{\Stwo}+\mathbf E^{\Vtwo}\).
The \(\Stwo\) term is distinct from the ordinary paraxial diffraction already
contained in the slow \(z\)-dependence of the leading scalar envelope.
The gradients do not follow the same naive ordering as the fields because a longitudinal carrier derivative contributes \(\partial_z\sim ik\), whereas
transverse derivatives act over the beam waist and scale as \(\nabla_\perp\sim w_0^{-1}\sim k\epsp\). Relative to the carrier-scale quadrupole amplitude \(kE_0\), the relevant gradient classes are
\begin{align}
\partial_z\mathbf E^{\Tzero}&\sim O(1),
&
\nabla_\perp\mathbf E^{\Tzero}&\sim O(\epsp),
\nonumber\\
\partial_z\mathbf E^{\Lone}&\sim O(\epsp),
&
\nabla_\perp\mathbf E^{\Lone}&\sim O(\epsp^2),
\nonumber\\
\partial_z\mathbf E^{\Stwo},
\ \partial_z\mathbf E^{\Vtwo}&\sim O(\epsp^2),
&
\nabla_\perp\mathbf E^{\Stwo},
\ \nabla_\perp\mathbf E^{\Vtwo}&\sim O(\epsp^3).
\label{eq:gradient_scaling_epsilon}
\end{align}
Thus, although the longitudinal field amplitude is first order in \(\epsp\),
its longitudinal gradient is of the same order as the transverse structured
gradient of the leading paraxial field. This is the basic reason longitudinal
optical structure can enter electric-quadrupole absorption at leading
structured-light order.

The truncation at second order is the minimal retained-order description that
includes the leading transverse field, the first longitudinal field required by
Maxwell transversality, the scalar nonparaxial envelope correction, and the
first transverse vectorial correction. Higher
orders give smaller corrections in the paraxial parameter and do not change the
leading transversality-locking relation
\begin{equation}
\nabla\cdot\mathbf E=0
\quad\Longrightarrow\quad
\nabla_\perp\cdot\mathbf E_\perp^{\Tzero}
+
\partial_zE_z^{\Lone}
=
0
\quad\Longrightarrow\quad
\partial_zE_z^{\Lone}
=
-\nabla_\perp\cdot\mathbf E_\perp^{\Tzero}.
\label{eq:transversality_locking}
\end{equation}
This relation is the central mechanism identified here.

\section{Vectorial Laguerre-Gaussian mode}\label{sec:field}

We now specialise the retained-order expansion to a fully vectorial
Laguerre-Gaussian mode. We treat the incident beam as a classical complex
spatial mode profile \(\mathbf E(\mathbf r)\), and write the scalar optical
mode profile as
\begin{align}
\psilp(\mathbf r)
&=
\fLG(r,z)e^{i(kz+\ell\phi)},
\label{eq:psi_def}
\end{align}
where
\begin{align}
\fLG(r,z)
&=
\sqrt{\frac{2p!}{\pi w_0^2(p+|\ell|)!}}
\frac{w_0}{w[z]}
\left(
\frac{\sqrt{2}r}{w[z]}
\right)^{|\ell|}
L_p^{|\ell|}
\left[
\frac{2r^2}{w^2[z]}
\right]
\nonumber\\
&\quad\times
\exp\left(-\frac{r^2}{w^2[z]}\right)
\exp\left[
i\left(
\frac{kr^2}{2R[z]}
-(2p+|\ell|+1)\zeta[z]
\right)
\right].
\label{eq:fLG}
\end{align}
Here \(w_0\) is the beam waist, \(w(z)\) is the beam radius, \(R(z)\) is the
wavefront radius of curvature, \(\zeta(z)\) is the Gouy phase, and
\(L_p^{|\ell|}\) is an associated Laguerre polynomial. The factor \(e^{ikz}\)
is included in \(\psilp\), while \(\fLG\) contains only the slowly varying
scalar diffraction structure and the transverse vortex phase is carried by
\(e^{i\ell\phi}\).

Beyond the leading paraxial mode, a consistent continuation through \(O(\epsp^2)\) requires corrections to both the scalar envelope and the
vectorial field structure. The scalar correction ensures that the scalar field satisfies the Helmholtz equation to the same order, while the Maxwell (vectorial) corrections also generate the first-order longitudinal and second-order cross-handed vectorial corrections introduced below. The scalar correction remains co-handed with the leading transverse field. We write
\begin{equation}
\psilp=e^{ikz}u_0,
\qquad
\psilpS=e^{ikz}u_2=\deltaS\,\psilp.
\label{eq:scalar_field_correction_def}
\end{equation}
Here \(u_0=\fLG e^{i\ell\phi}\) and \(u_2/u_0=O(\epsp^2)\).
The correction \(\psilpS\) is distinct from the ordinary paraxial
diffraction already contained in \(\fLG(r,z)\). In the notation of
Ref.~\cite{Alharbi2026}, \(\psi_2\) denotes the correction to the
generating scalar envelope, whereas \(\psilpS\) denotes the complete
co-handed electric-field correction. The latter also contains the
isotropic transverse-projection term generated by the relation between
the electric field and the vector potential. We adopt the
propagation-induced solution \(\psi_2=\psi_2^{(0)}\), as detailed in
Appendix~\ref{app:scalar}. The circular polarization basis is
\begin{align}
\mathbf e_\sigma
&=
\frac{1}{\sqrt{2}}
\left(
\hat{\mathbf x}+i\sigma\hat{\mathbf y}
\right),
&
\mathbf e_{-\sigma}
&=
\frac{1}{\sqrt{2}}
\left(
\hat{\mathbf x}-i\sigma\hat{\mathbf y}
\right),
\label{eq:circular_basis_main}
\end{align}
where \(\sigma=\pm1\) denotes the helicity. We use the dual-symmetrized Maxwell continuation of this paraxial input, in which the second-order vectorial correction is shared equally between the electric and magnetic fields. Retaining the leading transverse field, the first-order longitudinal field, the co-handed scalar correction, and the cross-handed vectorial correction gives \cite{Alharbi2026, forbes2026vortex}
\begin{align}
\mathbf{E}^{\leq2}_{\mathrm{LG}}(\mathbf r)
&=
\left(\psilp+\psilpS\right)\mathbf e_\sigma
+
\psilp\,\hat{\mathbf z}\,
\frac{i}{\sqrt{2}\,k}
\left(
\gamma-\frac{\sigma\ell}{r}
\right)e^{i\sigma\phi}
\nonumber\\
&\quad
+\frac{\psilp}{4k^2}
e^{i2\sigma\phi}
\left[
\gamma' + \gamma^2
-\frac{(1+2\sigma\ell)\gamma}{r}
+\frac{\ell^2+2\sigma\ell}{r^2}
\right]\mathbf e_{-\sigma}.
\label{eq:Efield_final_main}
\end{align}
The Cartesian reduction of the second-order terms is given in
Appendix~\ref{app:gradients}. Eq.~\eqref{eq:Efield_final_main} is the circular-polarization specialization of the dual-symmetrized field of Ref.~\cite{Alharbi2026}, truncated after the complete second-order transverse correction. The complete co-handed correction is the scalar
term \(\psilpS\mathbf e_\sigma\); the dual-symmetrized vectorial correction is
purely cross-handed. The radial quantity \(\gamma\) is the logarithmic radial derivative of the
scalar envelope,
\begin{align}
\gamma(\mathbf r)
&=
\frac{1}{\fLG}\frac{\partial \fLG}{\partial r}
=
\frac{|\ell|}{r}
-\frac{2r}{w^2(z)}
+\frac{ikr}{R(z)}
-\frac{4r}{w^2(z)}
\frac{L_{p-1}^{|\ell|+1}}{L_p^{|\ell|}},
\label{eq:gamma_main}
\end{align}
so that
\begin{equation}
\partial_r\fLG=\gamma\fLG,
\qquad
\gamma'=\partial_r\gamma.
\end{equation}
For \(p=0\), the term involving \(L_{p-1}^{|\ell|+1}\) is absent. All radial
combinations built from \(\gamma\) multiply the same external scalar mode
profile \(\psilp\). For longitudinal derivatives we separate the carrier derivative from scalar diffraction corrections by defining
\begin{align}
\chiz(\mathbf r)
&=
\frac{1}{\fLG}\frac{\partial\fLG}{\partial z},
\label{eq:chiz_def}
\\
\kappaz(\mathbf r)
&=
ik+\chiz(\mathbf r),
\label{eq:kappaz_def}
\end{align}
so that
\begin{equation}
\partial_z\psilp=\kappaz\psilp .
\label{eq:dzpsi_def}
\end{equation}
At the focal plane, for \(p=0\),
\begin{equation}
\chiz(r,0)
=
\frac{2i}{kw_0^2}
\left[
\frac{r^2}{w_0^2}-(|\ell|+1)
\right],
\label{eq:chiz_focus_p0}
\end{equation}
with \(|\ell|+1\) replaced by \(2p+|\ell|+1\) for general \(p\).
For the propagation-induced choice \(\psi_2=\psi_2^{(0)}\) specified in Appendix~\ref{app:scalar}, the complete focal-plane scalar correction obeys
\begin{equation}
ik\,\deltaS(r,0)=\chiz(r,0).
\label{eq:deltaS_chi_focus_relation}
\end{equation}
The derivation and explicit \(p=0\) form are given in Appendix~\ref{app:scalar}. In the retained gradient tensor, \(\psilpS\) contributes only through its carrier derivative \(ik\psilpS\mathbf e_\sigma\). The complete gradient bookkeeping, including the discarded higher-order derivatives, is given in Appendix~\ref{app:gradients}.

\section{Optical field-gradient tensor and transition channels}\label{sec:tensor}

The electric-quadrupole interaction couples the material quadrupole tensor to the optical field-gradient tensor. Since the electric quadrupole tensor is symmetric and traceless, only the symmetric-traceless part of the optical gradient contributes. We first form
\begin{equation}
G_{ij}
=
\frac{1}{2}
\left(
\partial_iE_j+\partial_jE_i
\right),
\label{eq:Gij_def}
\end{equation}
and project its traceless part onto a spherical rank-two basis. Because phase and sign conventions for spherical tensors vary, we state the convention explicitly:
\begin{align}
T^{(2)}_{\pm2}
&=
\frac{1}{2}
\left(
\partial_x E_x - \partial_y E_y
\right)
\pm
\frac{i}{2}
\left(
\partial_x E_y + \partial_y E_x
\right),
\label{eq:T2pm2}
\\
T^{(2)}_{\pm1}
&=
\mp
\frac{1}{2}
\left[
(\partial_x E_z + \partial_z E_x)
\pm i(\partial_y E_z + \partial_z E_y)
\right],
\label{eq:T2pm1}
\\
T^{(2)}_0
&=
\frac{1}{\sqrt6}
\left(
2\partial_zE_z-\partial_xE_x-\partial_yE_y
\right).
\label{eq:T20}
\end{align}
These are optical spherical tensor components. They should not be identified directly with molecular magnetic-sublevel changes. In the scalar quadrupole amplitude the optical and molecular tensors contract as $Q_q^{(2)}T_{-q}^{(2)}$. Since $Q_q^{(2)}$ connects molecular sublevels with $\Delta m=q$, a molecular transition with $\Delta m=q$ samples the optical component $T_{-q}^{(2)}$. To keep the notation readable, we define the transition-labelled optical amplitude
\begin{equation}
\Atrans_{\Delta m}^{(2)}
\equiv
T_{-\Delta m}^{(2)}.
\label{eq:A_transition_def}
\end{equation}
Thus
\begin{equation}
\Atrans_0^{(2)}=T_0^{(2)},
\qquad
\Atrans_{+1}^{(2)}=T_{-1}^{(2)},
\qquad
\Atrans_{+2}^{(2)}=T_{-2}^{(2)},
\end{equation}
with the corresponding sign reversal for negative $\Delta m$.

The explicit optical components obtained from Eqs.~\eqref{eq:T2pm2}-\eqref{eq:T20}
are as follows. For \(q=0\),
\begin{align}
T^{(2)}_0
=
-\frac{\sqrt{3}}{2}
\left(
\gamma-\frac{\ell\sigma}{r}
\right)
e^{i\sigma\phi}\psi_{\ell p}.
\label{eq:T20final}
\end{align}

This component contains both the transverse structured gradient of the leading
transverse field and the longitudinal gradient of the first-order longitudinal
field. For \(q=\pm2\),
\begin{equation}
T^{(2)}_{\pm2}
=
\frac{1}{2\sqrt{2}}
\left[
\gamma+\frac{\ell\sigma}{r}
\mp
\left(
\sigma\gamma+\frac{\ell}{r}
\right)
\right]
e^{-i\sigma\phi}\psi_{\ell p}.
\label{eq:T2pm2final}
\end{equation}
These \(q=\pm2\) components involve only transverse derivatives of transverse
fields at the retained order.

For $q=\pm1$, four retained contributions appear: the longitudinal derivative of the leading transverse field, the transverse derivatives of the first-order longitudinal field, the carrier derivative of the co-handed scalar correction, and the carrier derivative of the cross-handed vectorial correction,
\begin{equation}
T^{(2)}_{\pm1}
=
T^{(2)}_{\pm1}[\partial_zE^{\Tzero}]
+
T^{(2)}_{\pm1}[\nabla_\perp E^{\Lone}]
+
T^{(2)}_{\pm1}[\partial_zE^{\Stwo}]
+
T^{(2)}_{\pm1}[\partial_zE^{\Vtwo}].
\label{eq:Tpm1_split_main}
\end{equation}
After simplification,
\begin{align}
T^{(2)}_{\pm1}
=
\mp
\frac{1}{2\sqrt{2}}
(1\mp\sigma)
\left(ik+2\chiz+ik\deltaS\right)\psi_{\ell p}
\mp
\frac{3i}{8\sqrt{2}k}
(1\pm\sigma)Q_\sigma(r)
e^{i(\sigma\pm1)\phi}\psi_{\ell p},
\label{eq:Tpm1_final_simplified}
\end{align}
where
\begin{align}
Q_\sigma(r)
&=
\gamma'+\gamma^2-\frac{(1+2\sigma\ell)\gamma}{r}
+\frac{\ell^2+2\sigma\ell}{r^2}.
\label{eq:Qsigma_main}
\end{align}
The carrier-bearing component \(q=-\sigma\) contains three distinct
second-order terms. One \(\chiz\) comes from the slow longitudinal derivative
of the leading paraxial envelope, the second comes from the transverse
gradient of \(\mathbf E^{\Lone}\), and \(ik\deltaS\) is the carrier derivative
of the co-handed scalar correction. At the focal plane,
Eq.~\eqref{eq:deltaS_chi_focus_relation} reduces their sum to
\begin{equation}
ik+2\chiz+ik\deltaS=ik+3\chiz.
\label{eq:focal_complete_carrier}
\end{equation}
In the opposite component \(q=\sigma\), the transverse-gradient contribution \(\nabla_\perp\mathbf E^{\Lone}\) and the carrier derivative \(\partial_z\mathbf E^{\Vtwo}\) have the same optical structure, both being proportional to \(Q_\sigma\psi_{\ell p}/k\). They therefore populate the same quadrupole channel and add coherently, producing the nonzero second-order amplitude in Eq.~\eqref{eq:Tpm1_final_simplified}. Because this component has
no zeroth-order carrier amplitude with which to interfere, its contribution to the probability begins at \(O(\epsilon^4)\) and is omitted from the probability retained through \(O(\epsilon^2)\).
For helicity \(\sigma=+1\),
\begin{align}
T^{(2)}_{+1}
=
-\frac{3i}{4\sqrt{2}k}
Q_+(r)e^{i2\phi}\psi_{\ell p},
\qquad
T^{(2)}_{-1}
=
\frac{1}{\sqrt{2}}
\left(ik+2\chiz+ik\deltaS\right)\psi_{\ell p}.
\label{eq:Tpm1_sigma_plus_main}
\end{align}
For helicity \(\sigma=-1\),
\begin{align}
T^{(2)}_{+1}
=
-\frac{1}{\sqrt{2}}
\left(ik+2\chiz+ik\deltaS\right)\psi_{\ell p},
\qquad
T^{(2)}_{-1}
=
\frac{3i}{4\sqrt{2}k}
Q_-(r)e^{-i2\phi}\psi_{\ell p}.
\label{eq:Tpm1_sigma_minus_main}
\end{align}
The leading carrier-scale $q=\pm1$ response is therefore selected by helicity. The molecular transition label is obtained only after the sign reversal in Eq.~\eqref{eq:A_transition_def}.

\subsection{Transversality locking and the $\Delta m=0$ channel}

The $q=0$ optical component \eqref{eq:T20} provides the clearest manifestation of transversality locking because it contains the longitudinal gradient $\partial_zE_z$ directly. Since $q=0$ also corresponds to the transition-labelled amplitude $\mathcal A_0^{(2)}=T_0^{(2)}$, this is the component sampled by a molecular $\Delta m=0$ quadrupole transition. The point is not that longitudinal fields affect only this channel, but that in this channel
Maxwell transversality fixes their contribution directly. Maxwell transversality gives, to the retained order,
\begin{equation}
\nabla\cdot\mathbf E=0
\qquad\Rightarrow\qquad
\nabla_\perp\cdot\mathbf E_\perp^{\Tzero}
+
\partial_zE_z^{\Lone}
=0.
\end{equation}
The longitudinal field amplitude is paraxially small,
\begin{equation}
\frac{|E_z^{\Lone}|}{|\mathbf E_\perp^{\Tzero}|}
\sim
\frac{1}{kw_0},
\end{equation}
but its longitudinal derivative is locked to the transverse divergence so that at the retained order,
\begin{equation}
\partial_zE_z^{\Lone}
=
-\nabla_\perp\cdot\mathbf E_\perp^{\Tzero}.
\label{eq:longitudinal_gradient_lock}
\end{equation}
Substitution into Eq.~\eqref{eq:T20} gives
\begin{align}
T_0^{(2)}
&=
\frac{1}{\sqrt6}
\left(
2\partial_zE_z-\partial_xE_x-\partial_yE_y
\right)
\nonumber\\
&=
-\frac{3}{\sqrt6}
\nabla_\perp\cdot\mathbf E_\perp^{\Tzero}.
\label{eq:T0_transversality_locked}
\end{align}
A transverse-only treatment would instead give
\begin{equation}
T_{0,\mathrm{T0-only}}^{(2)}
=
-\frac{1}{\sqrt6}
\nabla_\perp\cdot\mathbf E_\perp^{\Tzero}.
\label{eq:T0_transverse_only}
\end{equation}

Thus the full retained optical $q=0$ amplitude, which is also the transition-labelled $\Atrans_0^{(2)}$ amplitude, is larger than the transverse-only result by a factor of three. This factor does not disappear as the waist is increased; what decreases with waist is the whole structured $q=0$ channel relative to the carrier-scale $q=\pm1$ quadrupole response. The component structure is summarised in Table~\ref{tab:selectionrules}.

\begin{table}[t]
\caption{Optical spherical components and the molecular transitions they sample. The sign reversal follows from the contraction $Q_q^{(2)}T_{-q}^{(2)}$.}
\label{tab:selectionrules}
\begin{tabular}{c c l}
\toprule
Optical component & Transition-labelled amplitude & Leading field-gradient content \\
\midrule
$T_0^{(2)}$ & $\Atrans_0^{(2)}$, $\Delta m=0$ &
\parbox[t]{8.0cm}{Contains $2\partial_zE_z-\partial_xE_x-\partial_yE_y$. The longitudinal gradient is locked to the transverse divergence and is essential.} \\[6pt]

$T_{\pm1}^{(2)}$ & $\Atrans_{\mp1}^{(2)}$, $\Delta m=\mp1$ &
\parbox[t]{8.0cm}{Dominated by $\partial_zE_x$ and $\partial_zE_y$. The carrier-bearing component contains $ik+2\chiz+ik\deltaS$: two terms arise from paraxial diffraction and $\nabla_\perp E_z^{\Lone}$, while the third is the co-handed scalar $\Stwo$ correction. At focus this becomes $ik+3\chiz$. The cross-handed $\Vtwo$ field contributes to the opposite optical component.} \\[6pt]

$T_{\pm2}^{(2)}$ & $\Atrans_{\mp2}^{(2)}$, $\Delta m=\mp2$ &
\parbox[t]{8.0cm}{Contains only transverse derivatives of transverse fields at the retained order. No longitudinal-gradient contribution appears.} \\
\bottomrule
\end{tabular}
\end{table}

\section{Absorption probability and retained-order rate}\label{sec:rate}

The preceding sections determine the optical tensor weights using a classical complex mode profile. To convert these weights into an absorption probability we now restore the molecular-QED normalization. In the Power-Zienau-Woolley representation, the interaction Hamiltonian contains \cite{craig1998molecular, salam2009molecular, andrews2018perspective}
\begin{align}
H_{\mathrm{int}}(\xi)
&=
-\mu_i(\xi)E_i(\mathbf R_\xi)
-Q_{ij}(\xi)\nabla_jE_i(\mathbf R_\xi)
-\cdots,
\label{eq:Hint_standard}
\end{align}
where $\mu_i$ is electric-dipole operator and $Q_{ij}$ is the electric-quadrupole operator. For a single incident mode, the photon-annihilation part of the electric-field operator may be written as
\begin{equation}
\hat{\mathbf E}^{(+)}(\mathbf r)
=
\Omega\,\mathbf E(\mathbf r)\,\hat a_{\ell p\sigma},
\label{eq:Eplus_restore}
\end{equation}
where $\mathbf E(\mathbf r)$ is the complex mode profile used above. Acting on an $n$-photon state gives $\langle n-1|\hat a_{\ell p\sigma}|n\rangle=\sqrt n$. The spatial tensor algebra is therefore unchanged; the rate acquires the common prefactor $n|\Omega|^2$.

The electric-quadrupole (E2) absorption amplitude may be written in Cartesian form as
\begin{align}
M_{fi}^{\mathrm{E2}}
&=
-\Omega\sqrt n\,
Q_{ij}^{m0}\nabla_jE_i(\mathbf R),
\label{eq:E2_cartesian_amplitude_restore}
\end{align}
or equivalently in spherical form as
\begin{align}
M_{fi}^{\mathrm{E2}}
&=
-\Omega\sqrt n
\sum_{q=-2}^{2}(-1)^qQ_q^{(2)}T_{-q}^{(2)}.
\label{eq:E2_spherical_amplitude}
\end{align}
Using Eq.~\eqref{eq:A_transition_def}, a resolved transition with $\Delta m=q$ has the optical amplitude $\Atrans_q^{(2)}=T_{-q}^{(2)}$, and hence a rate
\begin{align}
\Gamma_{\Delta m=q}^{\mathrm{E2}}
&\propto
\left|Q_q^{(2)}\right|^2
\left|\Atrans_q^{(2)}\right|^2
=
\left|Q_q^{(2)}\right|^2
\left|T_{-q}^{(2)}\right|^2.
\label{eq:channel_rate_weight}
\end{align}
For resolved magnetic sublevels with a chosen quantization axis, a transition with \(\Delta m=q\) samples the optical component \(T_{-q}^{(2)}\). If the magnetic sublevels are unresolved and the initial ensemble is incoherent, the
observed signal is an incoherent sum over the allowed \(\Delta m\) channels, weighted by their quadrupole line strengths. For a randomly oriented molecular
sample, such as a gas or liquid, the corresponding isotropic observable is obtained by rotationally averaging the molecular quadrupole tensor contracted with the optical field-gradient tensor. For an unpolarized atomic ensemble, the analogous isotropic limit is obtained by averaging and summing over magnetic
sublevels rather than by averaging over molecular orientations.

The perturbative expansion must be applied to the complete
field-gradient amplitude. Writing
\begin{equation}
M^{\mathrm{E2}}_{fi}
=
M^{(0)}_{fi}+M^{(1)}_{fi}+M^{(2)}_{fi}+\cdots ,
\end{equation}
a probability retained through $O(\epsp^2)$ contains
\begin{align}
\left|M^{\mathrm{E2}}_{fi}\right|^2_{\leq2}
={}&
\left|M^{(0)}_{fi}\right|^2
+2\operatorname{Re}\!\left[
M^{(0)}_{fi}M^{(1)*}_{fi}
\right]
+\left|M^{(1)}_{fi}\right|^2
\nonumber\\
&+
2\operatorname{Re}\!\left[
M^{(0)}_{fi}M^{(2)*}_{fi}
\right].
\end{align}
The order labels here refer to the complete field-gradient
amplitude, rather than directly to the order of the corresponding
field component. In particular,
$\nabla_\perp\mathbf E^{\Lone}=O(\epsp^2)$
relative to the carrier-scale quadrupole amplitude, so that
$|\nabla_\perp\mathbf E^{\Lone}|^2=O(\epsp^4)$
in the probability.

For an isotropically oriented sample, contraction with the
symmetric-traceless rotational average gives
\begin{equation}
\left\langle
\left|M^{\mathrm{E2}}_{fi}\right|^2
\right\rangle
=
\frac{n|\Omega|^2}{5}
Q^{m0}_{\lambda\mu}Q^{m0*}_{\lambda\mu}
\sum_{q=-2}^{2}\left|T^{(2)}_q\right|^2 .
\label{eq:isotropic-spherical-rate}
\end{equation}
The derivation is given in Appendix~\ref{app:isotropic-rate}.

Using Eqs.~\eqref{eq:T20final}-\eqref{eq:Tpm1_sigma_minus_main}, and retaining the probability through
$O(\epsp^2)$, gives
\begin{equation}
\left.
\sum_{q=-2}^{2}\left|T^{(2)}_q\right|^2
\right|_{\leq2}
=
\left[
\Kz
+\frac{3}{4}
\left|\gamma-\frac{\ell\sigma}{r}\right|^2
+\frac{1}{2}
\left|\gamma+\frac{\ell\sigma}{r}\right|^2
\right]
|f_{\mathrm{LG}}|^2 ,
\label{eq:retained-optical-sum}
\end{equation}
where
\begin{equation}
\Kz(r,z)
=
\frac{k^2}{2}
+2\operatorname{Re}\!\left[
ik\chi_z^*(r,z)
\right]
+k^2\operatorname{Re}\!\left[
\deltaS(r,z)
\right]
\label{eq:Kz_corrected}
\end{equation}
is the retained part of
\(\left|ik+2\chi_z+ik\deltaS\right|^2/2\). Terms quadratic in
\(\chi_z\) and \(\deltaS\), and their mutual interference, are fourth order
and are omitted. At the focal plane,
\begin{equation}
\Kz(r,0)
=
\frac{k^2}{2}
+3\operatorname{Re}\!\left[
ik\chi_z^*(r,0)
\right].
\label{eq:Kz_focus}
\end{equation}

The retained isotropic probability is therefore
\begin{align}
\left\langle
\left|M^{\mathrm{E2}}_{fi}\right|^2
\right\rangle_{\leq2}
={}&
\frac{n|\Omega|^2}{10}
Q^{m0}_{\lambda\mu}Q^{m0*}_{\lambda\mu}
\left[
2\Kz
+\frac{5}{2}|\gamma|^2
-\operatorname{Re}\!\left(
\frac{\ell\sigma}{r}\gamma^*
\right)
+\frac{5\ell^2}{2r^2}
\right]
|f_{\mathrm{LG}}|^2 .
\label{eq:retained-isotropic-rate}
\end{align}

\section{Numerical illustrations and physical interpretation}\label{sec:results}

We illustrate the analytical results using focal-plane line cuts and beam-waist maps. Unless stated otherwise, all numerical profiles in this section are evaluated at \(z=0\) for \(p=0\), with selected values of \(\ell\) and \(\sigma\). In this limit,
\begin{equation}
\gamma(r,0)=\frac{|\ell|}{r}-\frac{2r}{w_0^2},
\end{equation}
and
\begin{equation}
\kappaz(r,0)
=
ik+
\frac{2i}{kw_0^2}
\left[
\frac{r^2}{w_0^2}-(|\ell|+1)
\right].
\end{equation}
The carrier-bearing \(q=\pm1\) component contains the complete retained
combination
\begin{equation}
ik+2\chiz(r,0)+ik\deltaS(r,0)
=ik+3\chiz(r,0)
=
ik+
\frac{6i}{kw_0^2}
\left[
\frac{r^2}{w_0^2}-(|\ell|+1)
\right].
\end{equation}

\begin{figure}[!b]
    \centering
    \includegraphics[width=0.9\linewidth]{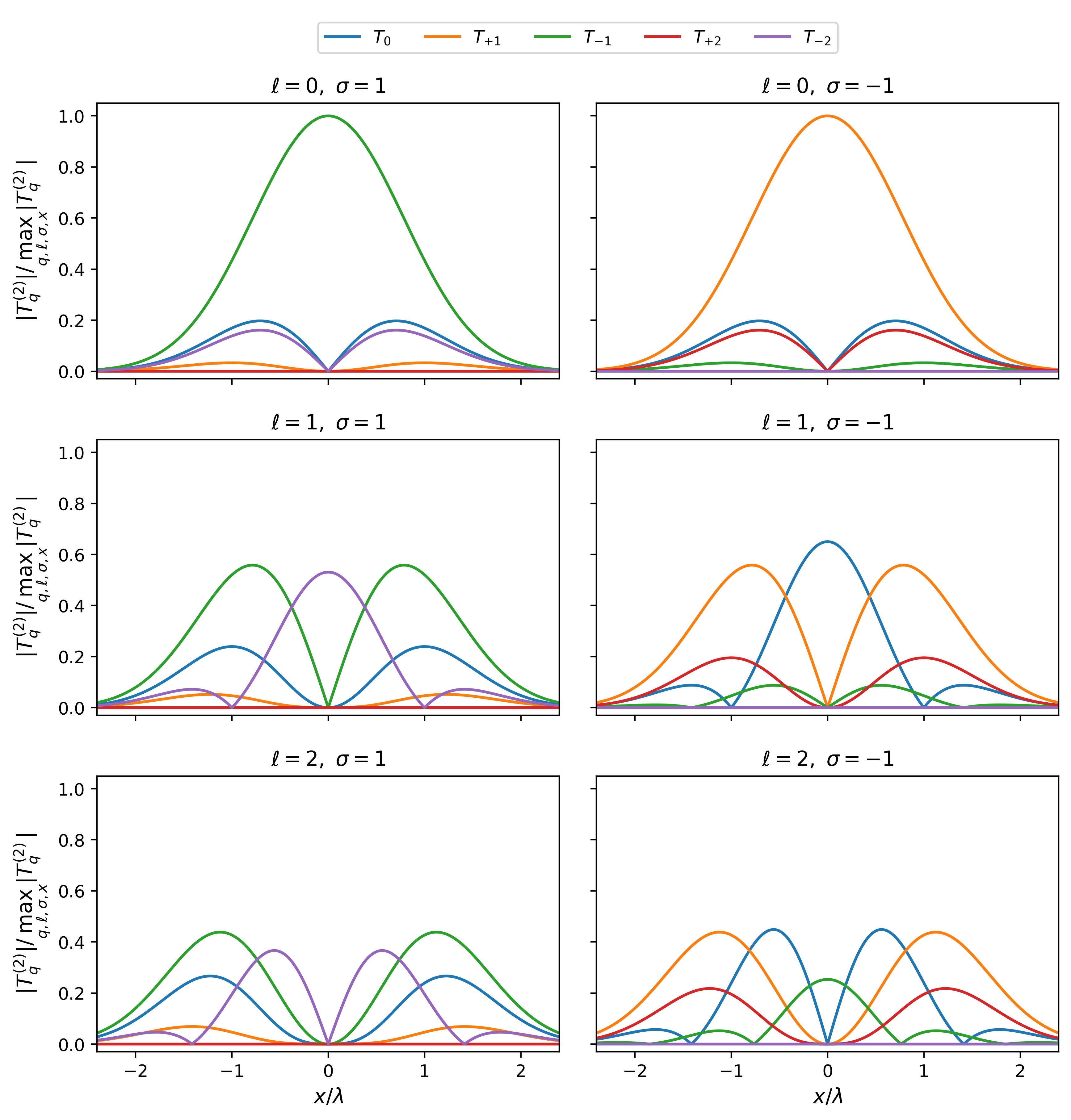}
    \caption{Radial dependence of the optical quadrupole tensor components \(T_q^{(2)}\) for \(p=0\) and selected values of \(\ell\) and \(\sigma\). The plotted index \(q\) is the optical tensor index. A molecular transition with label \(\Delta m\) samples the optical component \(T_{-\Delta m}^{(2)}\).}
    \label{fig:line_panels}
\end{figure}

The plotted quantities are optical tensor components \(T_q^{(2)}\). When the same data are interpreted as molecular transition amplitudes, the transition-labelled amplitude is
\(\Atrans_{\Delta m}^{(2)}=T_{-\Delta m}^{(2)}\). Figure~\ref{fig:line_panels} shows radial line cuts of the optical tensor components for representative values of \(\ell\) and \(\sigma\).

The optical \(q=\pm1\) components contain a carrier-scale term selected by helicity. For \(\sigma=+1\), \(T_{-1}^{(2)}\) is dominated at focus by \((ik+3\chiz)\psilp\) while \(T_{+1}^{(2)}\) is subleading; for \(\sigma=-1\), the roles reverse. This is a spin selection of the optical tensor component. The corresponding molecular transition labels are obtained by the sign reversal \(\Delta m=-q\).

The \(q=\pm2\) components show an analogous helicity dependence, but without longitudinal-field contributions. From Eq.~\eqref{eq:T2pm2final}, \(T_{+2}^{(2)}\) vanishes for \(\sigma=+1\) and \(T_{-2}^{(2)}\) vanishes for \(\sigma=-1\) within the retained theory.

\begin{figure}[!b]
    \centering
    \includegraphics[width=0.95\linewidth]{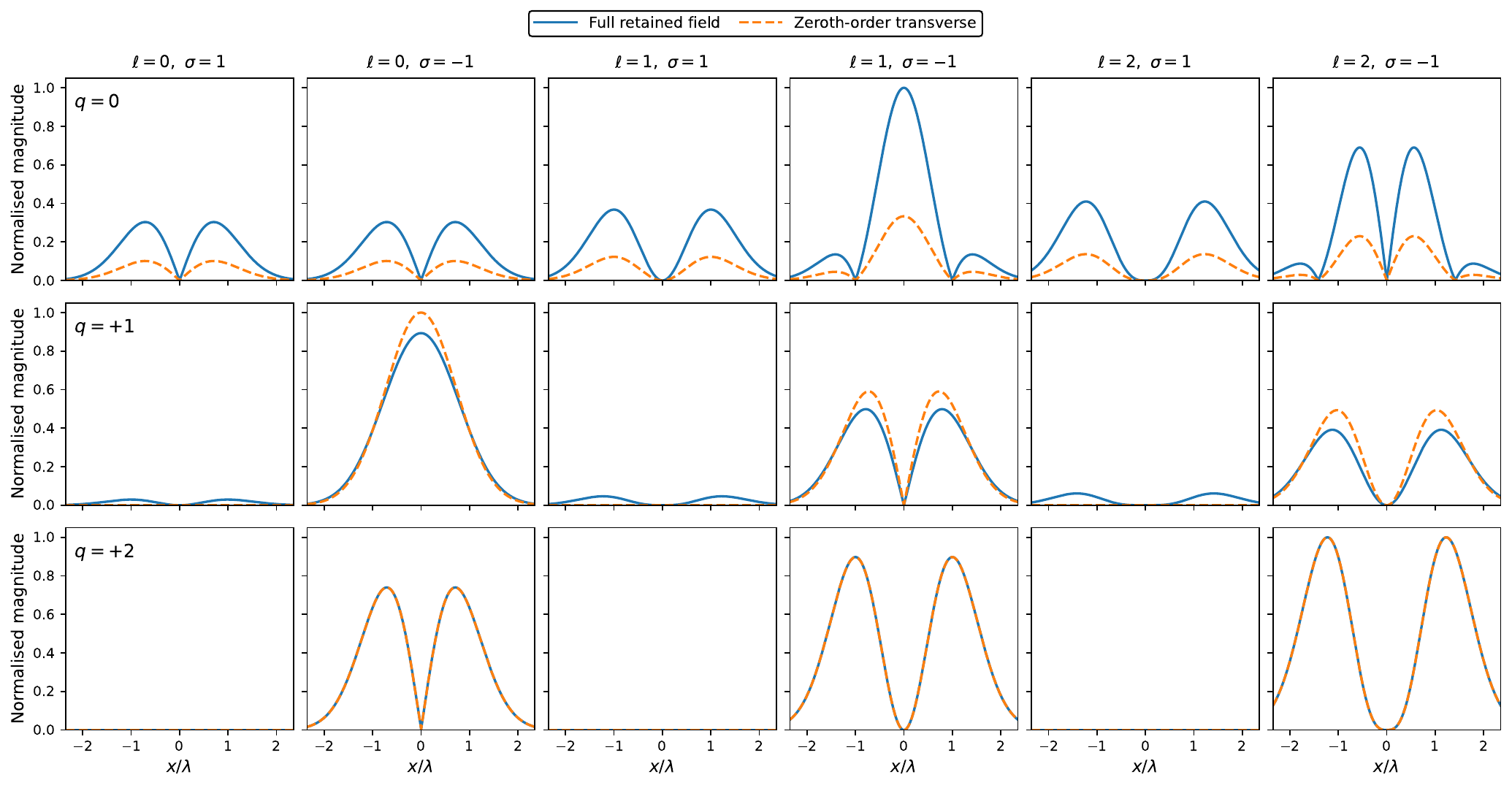}
    \caption{Comparison of the complete retained optical tensor components, including scalar and vectorial corrections, with the zeroth-order transverse-only approximation for $p=0$. Rows correspond to $T_0^{(2)}$, $T_{+1}^{(2)}$, and $T_{+2}^{(2)}$, not directly to molecular $\Delta m$ labels. Under Eq.~\eqref{eq:E2_spherical_amplitude}, these optical rows sample molecular transitions $\Delta m=0,-1,-2$, respectively. The panels are row-normalised to emphasise mechanism and spatial structure.}
    \label{fig:tq_full_vs_transverse}
\end{figure}

The \(q=0\) component behaves differently. It is not controlled by the carrier-scale longitudinal derivative of the transverse field. Instead it is controlled by the structured divergence of the transverse field and the longitudinal derivative of the longitudinal field. These two terms are tied by Eq.~\eqref{eq:transversality_locking}. Figure~\ref{fig:tq_full_vs_transverse} therefore shows that the transverse-only approximation misses a fixed part of the \(q=0\) amplitude. The factor of three in Eq.~\eqref{eq:T0_transversality_locked} is a consequence of Maxwell transversality.

\begin{figure}[!b]
    \centering
    \includegraphics[width=0.85\linewidth]{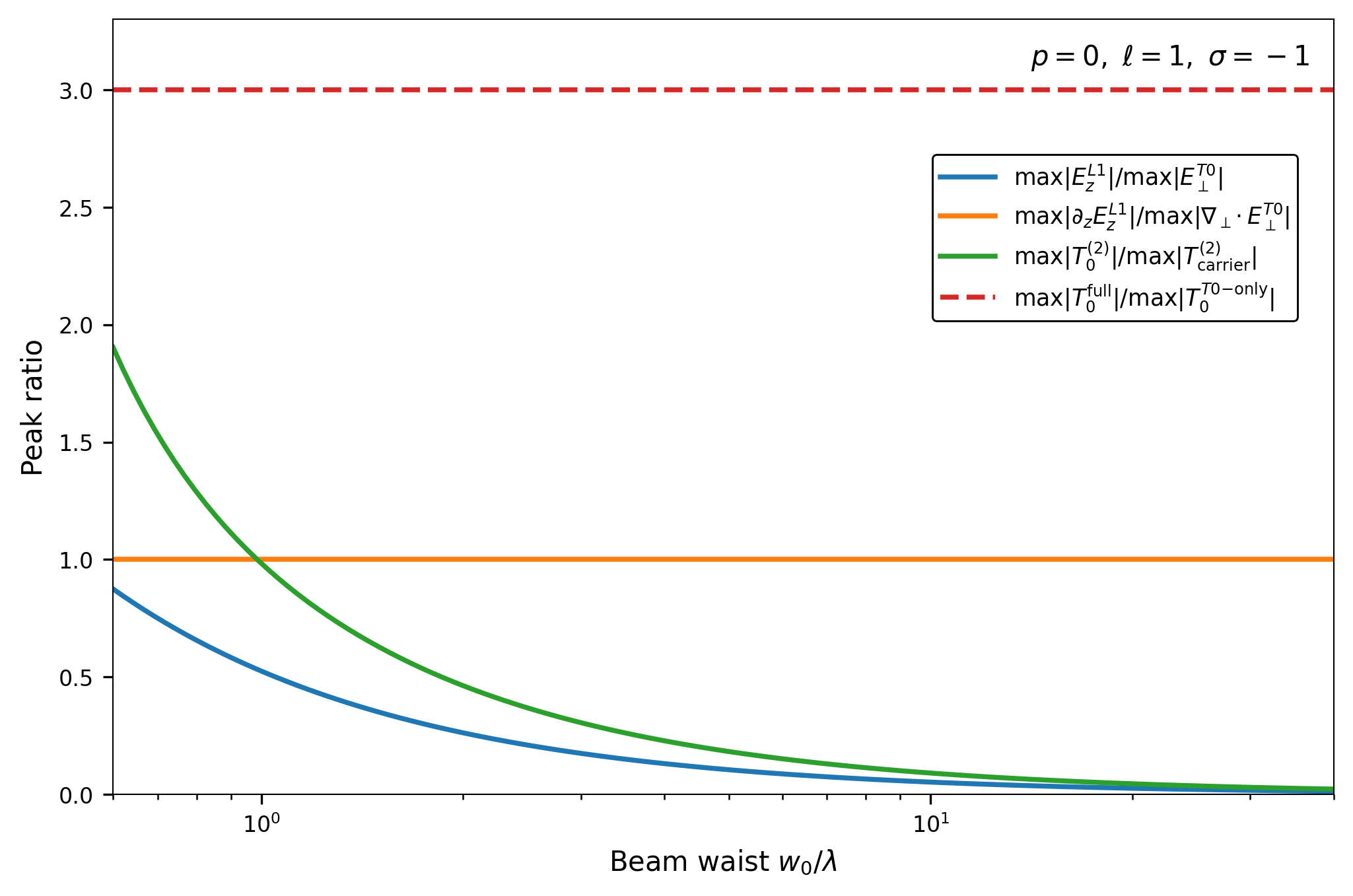}
    \caption{Beam-waist scaling of longitudinal-field and longitudinal-gradient effects for the optical \(q=0\) quadrupole component, shown for \(p=0\), \(\ell=1\), and \(\sigma=-1\). The horizontal axis is logarithmic in \(w_0/\lambda\), while the vertical axis is linear so that the peak ratios can be read directly. The longitudinal field amplitude, \(\max |E_z^{\Lone}|/\max |E_\perp^{\Tzero}|\), decreases with increasing waist, reflecting the usual paraxial suppression of \(E_z\). In contrast, the longitudinal gradient remains locked to the transverse divergence, giving \(\max |\partial_zE_z^{\Lone}|/\max |\nabla_\perp\cdot E_\perp^{\Tzero}|=1\) at the retained order. The full \(q=0\) amplitude remains a fixed factor of three larger than the transverse-only \(q=0\) result, while the structured \(q=0\) channel decreases relative to the carrier-scale quadrupole response as the beam waist increases.
}
    \label{fig:waist_scaling}
\end{figure}

\begin{figure}[!b]
    \centering
    {\includegraphics[width=0.85\linewidth]{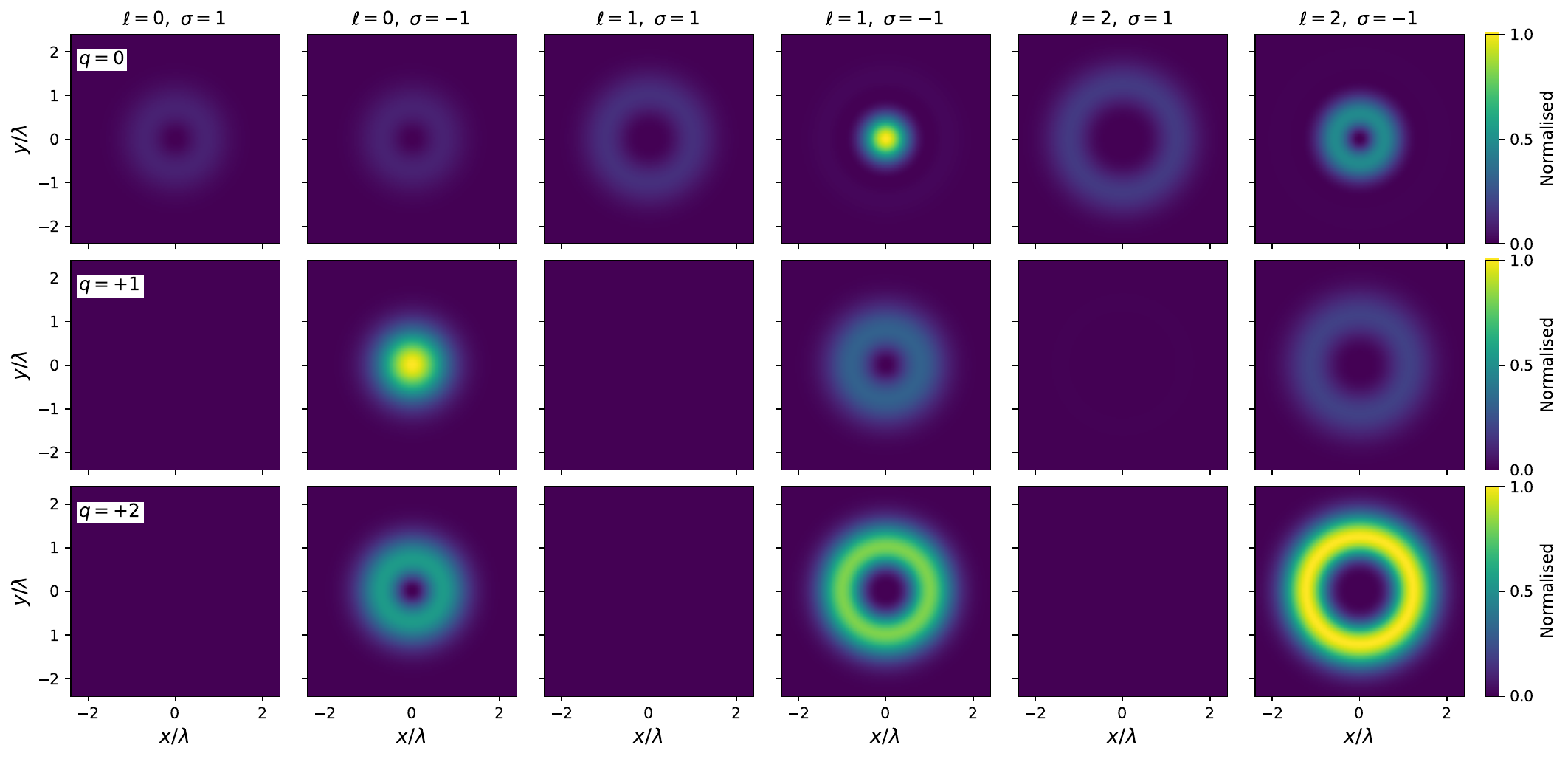}}
    \caption{Spatial distributions of the channel-resolved squared optical amplitudes in the beam-waist plane $z=0$ for $p=0$ and $w_0=\lambda$. The rows show $|T^{(2)}_0|^2$, $|T^{(2)}_{+1}|^2$, and $|T^{(2)}_{+2}|^2$, respectively. Under Eq.~\eqref{eq:A_transition_def}, these optical components sample molecular transitions $\Delta m=0,-1,-2$. Each row is normalized to the maximum value attained across that row. These diagnostic profiles are obtained by squaring the retained channel amplitudes and should not be identified with the consistently truncated isotropic probability of Eq.~\eqref{eq:retained-isotropic-rate}.
}
    \label{fig:rate_maps}
\end{figure} 

Orbital angular momentum enters through the combinations $\gamma-\ell\sigma/r$ and $\gamma+\ell\sigma/r$ that appear in the different tensor components. For $\ell=0$, the two helicities are mirror images and the response is primarily spin driven. For $\ell>0$, the spin and orbital terms may either cancel or reinforce one another. For example, at $\ell=1$ and $\sigma=+1$,
\begin{equation}
\gamma-\frac{\ell\sigma}{r}=-\frac{2r}{w_0^2},
\end{equation}
so the OAM contribution cancels from $T_0^{(2)}$. For $\ell=1$ and $\sigma=-1$,
\begin{equation}
\gamma-\frac{\ell\sigma}{r}=\frac{2}{r}-\frac{2r}{w_0^2},
\end{equation}
so the orbital and radial-gradient terms reinforce. This explains the strong helicity-dependent enhancement of $T_0^{(2)}$ and the allowed $T_{+2}^{(2)}$ component in the numerical plots.

Figure~\ref{fig:waist_scaling} separates three distinct notions of longitudinal importance. The ratio $\max|E_z^{\Lone}|/\max|\mathbf E_\perp^{\Tzero}|$ decreases approximately as $(kw_0)^{-1}$, which is the usual paraxial field-amplitude scaling. However, the ratio $\max|\partial_zE_z^{\Lone}|/\max|\nabla_\perp\cdot\mathbf E_\perp^{\Tzero}|$ remains unity at the retained order because of transversality. The structured $q=0$ channel nevertheless becomes smaller relative to the carrier-scale $q=\pm1$ response as the waist increases. Weak focusing therefore suppresses structured quadrupole channels relative to the dominant carrier-scale response, but it does not make the longitudinal part of the $q=0$ structured gradient dispensable.

Figure~\ref{fig:rate_maps} shows the corresponding two-dimensional
channel-resolved squared-amplitude profiles in the beam-waist plane. The rows display the optical components $q=0,+1,+2$, which sample molecular transitions $\Delta m=0,-1,-2$, respectively, according to Eq.~\eqref{eq:A_transition_def}. The $q=0$ profile shows the spatial structure of the transversality-locked longitudinal-gradient contribution. The $q=+1$ row contains the helicity-selected carrier response
for $\sigma=-1$ and the subleading vectorial response for $\sigma=+1$, while the $q=+2$ component is purely transverse at the retained order. These profiles are diagnostic squared magnitudes of the retained channel amplitudes; they are not separate terms in the second-order isotropic probability of Eq.~\eqref{eq:retained-isotropic-rate}.

These results show that the total angular momentum $J_z=\ell+\sigma$ alone is not a reliable predictor of the quadrupole response. The interaction is governed by how spin and orbital structure enter the individual optical gradient tensor components. Configurations with the same $J_z$ can therefore exhibit different quadrupole amplitudes and spatial profiles.

For a trapped atom whose magnetic sublevels are spectrally resolved, the transversality-locked relation is associated specifically with the transition channel $\Delta m=0$ when the atomic quantization axis is aligned with the beam propagation direction. However, this is not true when there is an angle $\theta$ between the two axes, as the beam-frame component \(T^{(2)}_{0}\) may contribute to several atomic transition channels, including $\Delta m=\pm2$, according to the Wigner rotation coefficients \(d^{(2)}_{q0}(\theta)\). The locking relation remains valid when the axes are tilted, but its identification with a single transition channel does not. In this case, the optical spherical tensor is rotated into the atomic frame.

In addition to coupling the atom's internal electronic states, a spatially structured field can coherently couple to its quantized center-of-mass motion \cite{stopp2022coherent}.
An electric-quadrupole amplitude depends on first spatial derivatives of the field, whereas first-order motional sidebands depend on second spatial derivatives \cite{verde2023trapped}. In the aligned \(\Delta m=0\) geometry, a sideband along direction \(\nu\) is therefore proportional to
$\partial_{\nu}T^{(2)}_{0}$,
and inherits the same transversality locking.

\clearpage
\section{Discussion and Conclusion}\label{sec:conclusion}

Previous work has established that longitudinal optical fields can be essential in electric-quadrupole excitation by structured light. In trapped-ion quadrupole excitation, the longitudinal component of a paraxial vortex beam was shown to be necessary for reproducing the observed selection-rule structure, particularly near the vortex axis where transverse-field models can fail \cite{quinteiro2017twisted}. This conclusion was subsequently extended beyond a single Laguerre-Gaussian construction to more general paraxial optical vortices interacting with atoms \cite{rosen2023importance}. Position-dependent quadrupole selection rules have also been derived with the longitudinal electric field included explicitly, showing that longitudinal contributions occur more broadly than the special antiparallel spin-orbit case and can affect Gaussian beams, parallel spin-orbit configurations, and channels that are inaccessible in a transverse-only description \cite{alharbi2023significance}. More recent studies have examined higher-order electric-field corrections in twisted-light quadrupole interactions \cite{alharbi2025role}, while general vectorial treatments of trapped atoms in structured light have shown that longitudinal fields and field gradients can strongly modify atomic transition matrix elements near the diffraction limit \cite{verde2023trapped}. Recent single-atom experiments using spatially structured light for narrow-line electric-quadrupole excitation further demonstrate that structured optical gradients are becoming practical tools for atomic control, cooling, and background-free imaging \cite{blodgett2025narrow}.

The present work builds on this literature but differs in emphasis and construction. Rather than treating the longitudinal field as an additional component whose importance must be assessed separately for each beam and transition, we identify the underlying order-by-order constraint imposed by Maxwell transversality. The first-order longitudinal field is smaller in amplitude than the leading transverse field, with $E_z^{\Lone}/E_\perp^{\Tzero}\sim(kw_0)^{-1}$ for a paraxial Laguerre-Gaussian beam. However, electric-quadrupole absorption is driven by field gradients, not by field amplitudes alone. The longitudinal derivative of the first-order longitudinal field recovers one power of the optical wavenumber and is locked to the transverse divergence of the leading transverse field. Thus $\partial_z E_z^{\Lone}$ and $\nabla_\perp E_\perp^{\Tzero}$ are contributions of the same structured-gradient order, and they must be retained together. The underlying locking relation is general to source-free electromagnetic fields; the role of structured light is to make the transverse divergence, and hence the locked longitudinal gradient, a controllable part of the quadrupole interaction. 

This transversality locking gives a simple organizing principle for longitudinal-field effects in structured-light quadrupole transitions. It explains why the longitudinal field can be paraxially small as a field amplitude while still making a leading contribution to the quadrupole tensor. In particular, for the optical \(q=0\) spherical component, Maxwell transversality gives \(\partial_zE_z^{\Lone}=-\nabla_\perp\cdot\mathbf E_\perp^{\Tzero}\) at the retained order. The full retained \(q=0\) amplitude is therefore not a focusing correction added on top of the transverse result: it is the vectorially complete form of the same structured-gradient channel. Weak focusing suppresses the channel as a whole, but it does not make the longitudinal part optional. 

The spherical tensor decomposition also shows that the effect is component selective. The optical $q=0$ component, which samples the molecular $\Delta m=0$ transition, depends on
$2\partial_zE_z-\partial_xE_x-\partial_yE_y$ and therefore requires the longitudinal gradient. The optical $q=\pm2$ components, which sample $\Delta m=\mp2$ transitions, are transverse at the retained order and contain no longitudinal-gradient contribution. The optical $q=\pm1$ components, which sample $\Delta m=\mp1$ transitions, are dominated by the carrier-scale longitudinal derivative of the transverse field. In the carrier-bearing component, second-order consistency gives \(ik+2\chiz+ik\deltaS\): the slow derivative of the leading paraxial envelope contributes once, the transverse gradient of the first-order longitudinal field supplies an equal \(\chiz\) term through the paraxial wave equation, and the carrier derivative of the co-handed scalar correction supplies \(ik\deltaS\). The cross-handed vectorial correction contributes only to the opposite optical component.

A second distinction of the present work is the consistent treatment of the retained-order absorption probability. The truncation at second order is chosen because it is the lowest order that contains the leading transverse field, the
first longitudinal field required by Maxwell transversality, and the first transverse vectorial correction. This is sufficient to capture the leading structured-gradient effect identified here, while keeping the paraxial expansion
controlled. Once the field is retained through the complete second-order transverse correction \(\mathbf E^{\Ttwo}=\mathbf E^{\Stwo}+\mathbf E^{\Vtwo}\), however, the probability cannot be obtained by simply squaring the truncated field and keeping all resulting terms. A probability retained through second order contains interference between zeroth- and second-order gradient amplitudes, but only after all contributions to each complete spherical component have been combined. Both \(\chiz\) and \(\deltaS\) in the carrier-bearing \(q=-\sigma\) component therefore survive through interference with the leading carrier amplitude and are included in \(\Kz\). By contrast, the \(Q_\sigma/k\) amplitude occupies the opposite component
\(q=\sigma\), which has no zeroth-order carrier term. Its squared magnitude is therefore of fourth order and is excluded. Retaining this contribution as part of a probability consistent through \(O(\epsp^4)\) would require the complete
field through fourth order, including the third order longitudinal and fourth-order transverse corrections, together with all associated interference terms through the same order.

Overall, these results show that longitudinal optical structure is not a removable higher-order detail in structured-light-induced electric-quadrupole absorption. Standard transverse-field approximations can remove leading-order structured-gradient contributions, especially in the \(\Delta m=0\) transition channel. The essential issue is not the size of the longitudinal field amplitude, but the Maxwell-locked structure of the optical field-gradient tensor. Future work should extend this analytical retained-order framework to higher orders in the Lax expansion, where additional vectorial corrections can be included consistently at the probability level. The same approach should also extend to higher-order multipole transitions, including electric-octupole transitions recently observed under vortex-beam excitation \cite{lange2022excitation}, magnetic-quadrupole transitions \cite{ji2024observation, salam2026magnetic}, and vector-structured (polarization-structured) light \cite{wang2020vectorial, svensson2025visualizing}. 

\appendix

\section{Scalar second-order correction}\label{app:scalar}

The leading LG envelope is
\begin{equation}
u_0(r,\phi,z)=\fLG(r,z)e^{i\ell\phi},
\qquad
\psilp=e^{ikz}u_0.
\end{equation}
To obtain the scalar field consistently through second order, write
\begin{equation}
u=u_0+u_2+O(\epsp^4),
\qquad
\frac{u_2}{u_0}=O(\epsp^2).
\end{equation}
The full carrier-bearing scalar field
\(\Psi=e^{ikz}u\) satisfies
\((\nabla^2+k^2)\Psi=0\). Factoring out the nonzero carrier gives the exact
envelope equation
\begin{equation}
\left(
\nabla_\perp^2+2ik\partial_z+\partial_z^2
\right)u=0.
\end{equation}
Separating successive paraxial orders yields
\begin{align}
\left(\nabla_\perp^2+2ik\partial_z\right)u_0&=0,
\label{eq:scalar_paraxial_u0}
\\
\left(\nabla_\perp^2+2ik\partial_z\right)u_2
&=-\partial_z^2u_0.
\label{eq:scalar_u2_equation}
\end{align}
Thus \(u_2\) cancels the next-order residual left by the paraxial solution
\(u_0\).

Equation~\eqref{eq:scalar_u2_equation} is a propagation equation and
requires specification of the scalar-envelope solution. To connect the
present notation with the dual-symmetrized construction of
Ref.~\cite{Alharbi2026}, it is important to distinguish the
scalar-envelope correction \(\psi_2\) from the complete co-handed
electric-field correction. In dimensionless variables, the latter is
\[
\psilpS
=
e^{i\zeta/f^2}f^2
\left[
\psi_2
+\frac{1}{2}
\left(\partial_\xi^2+\partial_\eta^2\right)\psi_0
\right].
\]
The coefficient \(1/2\) follows from the relation between the electric
field and the vector potential and is not an independently chosen
constant. We adopt the propagation-induced solution
\(\psi_2=\psi_2^{(0)}\), satisfying
\(\psi_2^{(0)}(\rho,0)=0\). Consequently, at the waist the complete
co-handed correction consists only of the transverse-projection term,
which in physical coordinates gives
\begin{equation}
u_2(r,\phi,0)
=
\frac{1}{2k^2}\nabla_\perp^2u_0(r,\phi,0).
\label{eq:scalar_u2_boundary}
\end{equation}
The choice of \(z=0\) specifies the reference plane; it does not restrict the
propagated field to the waist. A convenient solution satisfying
Eqs.~\eqref{eq:scalar_u2_equation} and \eqref{eq:scalar_u2_boundary} is
\begin{equation}
u_2
=
\frac{1}{2k^2}\nabla_\perp^2u_0
+
\frac{iz}{2k}\partial_z^2u_0.
\label{eq:scalar_u2_solution}
\end{equation}
The first term is the isotropic transverse-projection contribution, while the second is the propagation-induced scalar-envelope contribution. The corresponding electric-field term is
\[
\mathbf E^{\Stwo}
=\psilpS\mathbf e_\sigma,
\qquad
\psilpS=e^{ikz}u_2=\deltaS\psilp .
\]

At the waist, Eq.~\eqref{eq:scalar_u2_boundary} gives
\begin{align}
\deltaS(r,0)
&=
\frac{1}{2k^2}
\left(
\gamma'+\gamma^2+\frac{\gamma}{r}-\frac{\ell^2}{r^2}
\right)
\nonumber\\
&=
\frac{2}{k^2w_0^2}
\left[
\frac{r^2}{w_0^2}-(|\ell|+1)
\right]
\qquad (p=0).
\label{eq:deltaS_focus_p0}
\end{align}
Combining this result with the paraxial equation gives
\begin{equation}
ik\,\deltaS(r,0)=\chiz(r,0),
\end{equation}
which is the focal-plane identity used in
Eq.~\eqref{eq:deltaS_chi_focus_relation}.

In the dimensionless variables
\(\xi=x/w_0\), \(\eta=y/w_0\),
\(\zeta=z/(kw_0^2)\), and \(f=(kw_0)^{-1}\), the same correction is
\begin{equation}
\psilpS
=
e^{i\zeta/f^2}f^2
\left[
\psi_2^{(0)}
+\frac{1}{2}
\left(\partial_\xi^2+\partial_\eta^2\right)\psi_0
\right],
\qquad
\psi_2^{(0)}
=
\frac{i\zeta}{2}\partial_\zeta^2\psi_0
=
-\frac{i\zeta}{8}\nabla_\perp^4\psi_0.
\label{eq:app_scalar_dimensionless}
\end{equation}

Thus \(\psi_2^{(0)}\) is the propagation-induced correction to the generating scalar envelope, while the complete co-handed electric-field correction is the full expression in square brackets. Alternative admissible choices of \(\psi_2\) correspond to different boundary or
asymptotic definitions of the beam. They modify the prescription-dependent scalar contribution to the carrier-bearing \(q=-\sigma\) channel, but do not affect the transversality-locked \(q=0\) result or the retained \(q=\pm2\) components.

At the retained order its gradient contributes only through the carrier
derivative,
\begin{equation}
\nabla_jE_i^{\Stwo}
\approx
ik\,\hat{\mathbf z}_j(\mathbf e_\sigma)_i\psilpS.
\label{eq:app_gradS2_longitudinal_only}
\end{equation}
Transverse derivatives of \(\mathbf E^{\Stwo}\) are third order relative to
the carrier-scale gradient, while slow longitudinal derivatives of
\(\psilpS\) are fourth order, and both are omitted.

\section{Field gradients and asymptotic ordering}\label{app:gradients}

For a general input Jones vector
\(\alpha\hat{\mathbf x}+\beta\hat{\mathbf y}\), the
complete second-order transverse electric correction in physical coordinates
is the sum of scalar and vectorial parts,
\begin{align}
E_x^{\Stwo+\Vtwo}
&=
\alpha\psilpS
+
\frac{1}{4k^2}
\left[
\alpha(\partial_x^2-\partial_y^2)
+2\beta\partial_x\partial_y
\right]\psilp,
\label{eq:app_dual_Ex}
\\
E_y^{\Stwo+\Vtwo}
&=
\beta\psilpS
+
\frac{1}{4k^2}
\left[
2\alpha\partial_x\partial_y
-\beta(\partial_x^2-\partial_y^2)
\right]\psilp.
\label{eq:app_dual_Ey}
\end{align}
Substitution of Eq.~\eqref{eq:app_scalar_dimensionless} into
Eqs.~\eqref{eq:app_dual_Ex} and \eqref{eq:app_dual_Ey} gives the Cartesian
coefficients \(3/4\), \(1/4\), and \(1/2\) of the complete
dual-symmetrized field directly.
For circular polarization,
\(\alpha=1/\sqrt2\) and \(\beta=i\sigma/\sqrt2\). Direct substitution gives
\begin{align}
\mathbf E^{\Stwo+\Vtwo}
&=
\psilpS\mathbf e_\sigma
+
\frac{1}{4k^2}
\left(\partial_x+i\sigma\partial_y\right)^2
\psilp\,\mathbf e_{-\sigma}
\nonumber\\
&=
\psilpS\mathbf e_\sigma
+
\frac{\psilp}{4k^2}
Q_\sigma(r)e^{i2\sigma\phi}\mathbf e_{-\sigma},
\label{eq:app_dual_T2_circular}
\end{align}
where \(Q_\sigma\) is defined below. Thus the co-handed term is entirely the
scalar correction, while the vectorial part is entirely cross-handed with
coefficient \(1/(4k^2)\), exactly as in
Eq.~\eqref{eq:Efield_final_main}.

The gradient of the leading transverse field is
\begin{align}
\nabla_jE^{\Tzero}_i(\mathbf r)
&=
(\mathbf e_\sigma)_i
\left(
\gamma\hat{\mathbf r}
+
\frac{i\ell}{r}\hat{\boldsymbol\phi}
+\kappaz\hat{\mathbf z}
\right)_j
\psilp .
\label{eq:app_gradT0}
\end{align}
In Eq.~\eqref{eq:app_gradT0}, $\kappaz$ is required only in the leading transverse carrier contribution. In lower-order field components, replacing $\partial_z\psilp$ by $ik\psilp$ is sufficient at the retained order.

The gradient of the longitudinal component is most compactly written by defining
\begin{equation}
B(r)\equiv \gamma-\frac{\ell\sigma}{r}.
\end{equation}
Then
\begin{align}
\nabla_jE^{\Lone}_i(\mathbf r)
&=
\frac{1}{\sqrt2}\hat{\mathbf z}_i\frac{i}{k}e^{i\sigma\phi}
\Biggl(
\hat{\mathbf r}_j\left[B'+\gamma B\right]
+
\hat{\boldsymbol\phi}_j\frac{i(\ell+\sigma)}{r}B
+ik\hat{\mathbf z}_j B
\Biggr)
\psilp
\nonumber\\
&=
\frac{1}{\sqrt2}\hat{\mathbf z}_i e^{i\sigma\phi}
\Biggl(
\frac{i}{k}\hat{\mathbf r}_j
\left[
\gamma'+\gamma^2
-\frac{\ell\sigma}{r}\gamma
+\frac{\ell\sigma}{r^2}
\right]
-
\hat{\boldsymbol\phi}_j
\frac{\ell+\sigma}{kr}
\left[
\gamma-\frac{\ell\sigma}{r}
\right]
-
\hat{\mathbf z}_j
\left[
\gamma-\frac{\ell\sigma}{r}
\right]
\Biggr)
\psilp .
\label{eq:app_gradL1}
\end{align}
The omitted replacement $ik\rightarrow\kappaz$ in the final longitudinal derivative of $E^{\Lone}$ would generate an amplitude correction of order $\epsp^3$ and is not retained.

Define
\begin{align}
Q_\sigma(r)
&=
\gamma'+\gamma^2-\frac{(1+2\sigma\ell)\gamma}{r}
+\frac{\ell^2+2\sigma\ell}{r^2},
\label{eq:app_Qsigma}
\\
\Rco(r)
&=
-\gamma'-\gamma^2-\frac{\gamma}{r}+\frac{\ell^2}{r^2}.
\label{eq:app_Rco}
\end{align}
The paraxial equation for the scalar LG envelope gives
\begin{equation}
\chiz
=
-\frac{i}{2k}\Rco .
\label{eq:app_chi_R_relation}
\end{equation}
For the propagation-induced choice
\(\psi_2=\psi_2^{(0)}\), the resulting complete focal-plane scalar correction gives
\begin{equation}
\deltaS(r,0)
=
-\frac{1}{2k^2}\Rco(r,0),
\qquad
ik\deltaS(r,0)=\chiz(r,0).
\label{eq:app_delta_R_relation}
\end{equation}
The second-order transverse vectorial gradient may be written compactly as
\begin{align}
\nabla_jE_i^{\Vtwo}(\mathbf r)
&=
\frac{1}{4k^2}
\Biggl(
(\mathbf e_{-\sigma})_i
\left[
\mathcal D_{\ell+2\sigma}(Q_\sigma)
\right]_j
e^{i2\sigma\phi}
\Biggr)
\psilp,
\label{eq:app_gradT2_compact}
\end{align}
where
\begin{align}
\mathcal D_m[F]
=
\hat{\mathbf r}\left(F'+\gamma F\right)
+
\hat{\boldsymbol\phi}\frac{im}{r}F
+ik\hat{\mathbf z}F.
\label{eq:app_Dm}
\end{align}
Retaining only the dominant longitudinal derivative gives
\begin{align}
\nabla_jE_i^{\Vtwo}(\mathbf r)
\approx
ik\,\hat{\mathbf z}_j\,E_i^{\Vtwo}(\mathbf r).
\label{eq:app_gradT2_longitudinal_only}
\end{align}
The slow-envelope derivatives of both second-order fields are fourth order relative to \(kE_0\) and are omitted.

\section{Orientationally averaged absorption probability}\label{app:isotropic-rate}

For an isotropically oriented symmetric-traceless
electric-quadrupole transition tensor,
\begin{equation}
\left\langle
Q^{m0}_{ij}Q^{m0*}_{kl}
\right\rangle
=
\frac{
Q^{m0}_{\lambda\mu}Q^{m0*}_{\lambda\mu}
}{10}
\left(
\delta_{ik}\delta_{jl}
+\delta_{il}\delta_{jk}
-\frac{2}{3}\delta_{ij}\delta_{kl}
\right).
\label{eq:quadrupole-rotational-average}
\end{equation}
The trace term vanishes after contraction with the optical
gradient in a source-free homogeneous region because
$\partial_iE_i=0$. It follows that
\begin{align}
\left\langle
\left|M^{\mathrm{E2}}_{fi}\right|^2
\right\rangle
={}&
\frac{n|\Omega|^2}{10}
Q^{m0}_{\lambda\mu}Q^{m0*}_{\lambda\mu}
\nonumber\\
&\times
\left[
(\partial_jE_i)(\partial_jE_i)^*
+
(\partial_jE_i)(\partial_iE_j)^*
\right].
\end{align}

Writing
\begin{equation}
G_{ij}
=
\frac{1}{2}
\left(
\partial_iE_j+\partial_jE_i
\right),
\end{equation}
the optical contraction becomes
\begin{equation}
(\partial_jE_i)(\partial_jE_i)^*
+
(\partial_jE_i)(\partial_iE_j)^*
=
2G_{ij}G_{ij}^*
=
2\sum_{q=-2}^{2}|T^{(2)}_q|^2 .
\end{equation}
Therefore
\begin{equation}
\left\langle
\left|M^{\mathrm{E2}}_{fi}\right|^2
\right\rangle
=
\frac{n|\Omega|^2}{5}
Q^{m0}_{\lambda\mu}Q^{m0*}_{\lambda\mu}
\sum_{q=-2}^{2}|T^{(2)}_q|^2 .
\end{equation}

For fixed helicity, Eqs.~\eqref{eq:T20final}-\eqref{eq:Tpm1_sigma_minus_main} give
\begin{equation}
\left.
\sum_{q=-2}^{2}|T^{(2)}_q|^2
\right|_{\leq2}
=
\left[
\Kz
+\frac{3}{4}
\left|\gamma-\frac{\ell\sigma}{r}\right|^2
+\frac{1}{2}
\left|\gamma+\frac{\ell\sigma}{r}\right|^2
\right]
|f_{\mathrm{LG}}|^2 .
\end{equation}

The $Q_\sigma/k$ term in the subleading $q=\sigma$
component is second order in the amplitude and therefore
contributes only at fourth order when squared. Similarly,
$\nabla_\perp\mathbf E^{\Lone}$ is second order in
the complete field-gradient amplitude. Its same-channel part
interferes with the carrier and is retained through the second
\(\chiz\) contribution in \(\Kz\), whereas its squared contribution is fourth
order and is excluded. The co-handed
\(\partial_z\mathbf E^{\Stwo}\) term also interferes with the carrier and
produces the \(k^2\operatorname{Re}\deltaS\) term in \(\Kz\). The
cross-handed \(\partial_z\mathbf E^{\Vtwo}\) term occupies \(q=\sigma\) and
therefore has no carrier with which to interfere.
\section{Spherical-tensor decomposition and channel-resolved derivations}\label{app:spherical}

Using
\begin{equation}
\partial_x
=
\cos\phi\,\partial_r
-\frac{1}{r}\sin\phi\,\partial_\phi,
\qquad
\partial_y
=
\sin\phi\,\partial_r
+\frac{1}{r}\cos\phi\,\partial_\phi,
\label{eq:app_cartpolar}
\end{equation}
and the normalized circular basis,
\begin{equation}
\mathbf e_\sigma
=
\frac{1}{\sqrt2}
\left(
\hat{\mathbf x}+i\sigma\hat{\mathbf y}
\right),
\end{equation}
one has
\begin{equation}
\nabla_\perp\cdot\mathbf E_\perp^{\Tzero}
=
\frac{1}{\sqrt2}
\left(
\gamma-\frac{\ell\sigma}{r}
\right)
e^{i\sigma\phi}\psilp .
\end{equation}
At the retained order,
\begin{equation}
\partial_zE_z^{\Lone}
=
-\frac{1}{\sqrt2}
\left(
\gamma-\frac{\ell\sigma}{r}
\right)
e^{i\sigma\phi}\psilp .
\end{equation}
Therefore
\begin{align}
T^{(2)}_0
&=
\frac{1}{\sqrt6}
\left(
2\partial_zE_z-\partial_xE_x-\partial_yE_y
\right)
\nonumber\\
&=
-\frac{3}{\sqrt{12}}
\left(
\gamma-\frac{\ell\sigma}{r}
\right)
e^{i\sigma\phi}\psilp
\nonumber\\
&=
-\frac{\sqrt3}{2}
\left(
\gamma-\frac{\ell\sigma}{r}
\right)
e^{i\sigma\phi}\psilp .
\label{eq:app_T20final}
\end{align}
The replacement $ik\rightarrow\kappaz$ in $\partial_zE_z^{\Lone}$ would enter
at third order in the amplitude and is not retained.

Similarly, the $q=\pm2$ component becomes
\begin{equation}
T^{(2)}_{\pm2}
=
\frac{1}{2\sqrt2}
\left[
\gamma+\frac{\ell\sigma}{r}
\mp
\left(
\sigma\gamma+\frac{\ell}{r}
\right)
\right]
e^{-i\sigma\phi}\psilp .
\label{eq:app_T2pm2final}
\end{equation}

For the $q=\pm1$ channels, write
\begin{equation}
B
\equiv
\gamma-\frac{\ell\sigma}{r}.
\label{eq:app_B_def}
\end{equation}
The relevant Cartesian circular combinations are
\begin{align}
E_x+iE_y
&=
\frac{1-\sigma}{\sqrt2}\left(\psilp+\psilpS\right)
+
\frac{1+\sigma}{4\sqrt2 k^2}
Q_\sigma e^{i2\sigma\phi}\psilp,
\label{eq:app_Ex_plus_iEy}
\\
E_x-iE_y
&=
\frac{1+\sigma}{\sqrt2}\left(\psilp+\psilpS\right)
+
\frac{1-\sigma}{4\sqrt2 k^2}
Q_\sigma e^{i2\sigma\phi}\psilp,
\label{eq:app_Ex_minus_iEy}
\\
E_z
&=
\frac{i}{\sqrt2 k}
B e^{i\sigma\phi}\psilp .
\label{eq:app_field_components_pm1}
\end{align}

The longitudinal gradient of the zeroth-order transverse field gives
\begin{align}
T^{(2)}_{+1}[\partial_zE^{\Tzero}]
&=
-\frac12\partial_z(E_x^{\Tzero}+iE_y^{\Tzero})
=
-\frac{1}{2\sqrt2}(1-\sigma)\kappaz\psilp,
\label{eq:app_Tplus1_T0}
\\
T^{(2)}_{-1}[\partial_zE^{\Tzero}]
&=
+\frac12\partial_z(E_x^{\Tzero}-iE_y^{\Tzero})
=
+\frac{1}{2\sqrt2}(1+\sigma)\kappaz\psilp .
\label{eq:app_Tminus1_T0}
\end{align}

The transverse gradients of the first-order longitudinal field give
\begin{align}
T^{(2)}_{+1}[\nabla_\perp E^{\Lone}]
&=
-\frac12(\partial_x+i\partial_y)E_z
\nonumber\\
&=
-\frac{i}{2\sqrt2 k}
\left[
B'+\gamma B-\frac{\ell+\sigma}{r}B
\right]
e^{i(\sigma+1)\phi}\psilp,
\label{eq:app_Tplus1_L1}
\\
T^{(2)}_{-1}[\nabla_\perp E^{\Lone}]
&=
+\frac12(\partial_x-i\partial_y)E_z
\nonumber\\
&=
+\frac{i}{2\sqrt2 k}
\left[
B'+\gamma B+\frac{\ell+\sigma}{r}B
\right]
e^{i(\sigma-1)\phi}\psilp .
\label{eq:app_Tminus1_L1}
\end{align}
Using the helicity values \(\sigma=\pm1\), Eqs.~\eqref{eq:app_Tplus1_L1}
and \eqref{eq:app_Tminus1_L1} may equivalently be written as
\begin{align}
T^{(2)}_{+1}[\nabla_\perp E^{\Lone}]
&=
\frac{i}{4\sqrt2 k}(1-\sigma)\Rco\,\psilp
-\frac{i}{4\sqrt2 k}(1+\sigma)
Q_\sigma e^{i(\sigma+1)\phi}\psilp,
\label{eq:app_Tplus1_L1_RQ}
\\
T^{(2)}_{-1}[\nabla_\perp E^{\Lone}]
&=
-\frac{i}{4\sqrt2 k}(1+\sigma)\Rco\,\psilp
+\frac{i}{4\sqrt2 k}(1-\sigma)
Q_\sigma e^{i(\sigma-1)\phi}\psilp.
\label{eq:app_Tminus1_L1_RQ}
\end{align}

The carrier derivative of the co-handed scalar correction gives
\begin{align}
T^{(2)}_{+1}[\partial_zE^{\Stwo}]
&=
-\frac{i k}{2\sqrt2}(1-\sigma)\psilpS,
\label{eq:app_Tplus1_S2}
\\
T^{(2)}_{-1}[\partial_zE^{\Stwo}]
&=
+\frac{i k}{2\sqrt2}(1+\sigma)\psilpS.
\label{eq:app_Tminus1_S2}
\end{align}

Retaining only the carrier contribution in the cross-handed vectorial field
gives
\begin{align}
T^{(2)}_{+1}[\partial_zE^{\Vtwo}]
&=
-\frac12\partial_z(E_x^{\Vtwo}+iE_y^{\Vtwo})
\nonumber\\
&=
-\frac{i}{8\sqrt2 k}
(1+\sigma)Q_\sigma e^{i2\sigma\phi}\psilp,
\label{eq:app_Tplus1_T2}
\\
T^{(2)}_{-1}[\partial_zE^{\Vtwo}]
&=
+\frac12\partial_z(E_x^{\Vtwo}-iE_y^{\Vtwo})
\nonumber\\
&=
+\frac{i}{8\sqrt2 k}
(1-\sigma)Q_\sigma e^{i2\sigma\phi}\psilp .
\label{eq:app_Tminus1_T2}
\end{align}

Using Eq.~\eqref{eq:app_chi_R_relation},
\begin{equation}
\kappaz-\frac{i}{2k}\Rco+ik\deltaS
=
ik+2\chiz+ik\deltaS .
\label{eq:app_carrier_combination}
\end{equation}
At the focal plane, Eq.~\eqref{eq:app_delta_R_relation} reduces this
combination to \(ik+3\chiz\). Combining all four retained contributions gives
\begin{align}
T^{(2)}_{+1}
&=
-\frac{1}{2\sqrt2}(1-\sigma)
\left(ik+2\chiz+ik\deltaS\right)\psilp
-\frac{3i}{8\sqrt2 k}
(1+\sigma)Q_\sigma e^{i(\sigma+1)\phi}\psilp,
\label{eq:app_Tplus1_final_corrected}
\\
T^{(2)}_{-1}
&=
+\frac{1}{2\sqrt2}(1+\sigma)
\left(ik+2\chiz+ik\deltaS\right)\psilp
+\frac{3i}{8\sqrt2 k}
(1-\sigma)Q_\sigma e^{i(\sigma-1)\phi}\psilp .
\label{eq:app_Tminus1_final_corrected}
\end{align}
These two equations are the detailed versions of the compact result quoted in
the main text as Eq.~\eqref{eq:Tpm1_final_simplified}.

\bibliography{references}

\end{document}